\documentclass[aps,prd,twocolumn,
superscriptaddress,preprintnumbers,
showpacs,nofootinbib,nobibnotes]{revtex4}
\usepackage{amsfonts,amssymb,amsmath}
\usepackage{graphicx}

\newcommand{\Valencia}{Departamento de F\'\i sica Te\'orica
and IFIC, Centro Mixto, Universidad de Valencia--CSIC \\
E-46100, Burjassot, Valencia, Spain}
\newcommand{\Dubna}{Bogoliubov Laboratory of Theoretical Physics,
Joint Institute for Nuclear Research \\
Dubna, 141980, Russian Federation}

\newcommand{\ind}[2]{^{#1}_{\text{\scriptsize #2}}}
\newcommand{\al}[2]{\alpha\ind{#1}{#2}}

\newcommand{\hal}[2]{\widehat{\alpha}\ind{#1}{#2}}
\newcommand{\ro}[1]{\rho^{(#1)}}
\newcommand{\Ro}[2]{{\cal R}^{(#1)}(#2)}
\newcommand{\RTau}[1]{R_{\tau,\text{\tiny #1}}}
\newcommand{\dQCD}[1]{\delta^{#1}_{\text{\tiny QCD}}}
\newcommand{\pslush}{p\hspace{-0.16cm}/}

\newcommand{\Vud}{V_{\text{\scriptsize ud}}}
\newcommand{\Sew}{S_{\text{\tiny EW}}}
\newcommand{\fpb}{\frac{4 \pi}{\bz}}
\newcommand{\Nc}{N_{\text{\scriptsize c}}}
\newcommand{\nf}{n_{\text{\scriptsize f}}}
\newcommand{\bz}{\beta_0}
\newcommand{\mpi}{m_{\pi}}
\newcommand{\MTau}{M_{\tau}}
\newcommand{\alc}{\alpha_{\text{\tiny C}}}
\newcommand{\KL}{K\"all\'en--Lehmann }
\newcommand{\ADf}{Adler $D$~function}
\newcommand{\JLD}{Jost--Lehmann--Dyson }
\newcommand{\SD}{Schwinger--Dyson }

\newcommand{\epem}{e^{+}e^{-}}
\newcommand{\cor}{{\cal O}\!}

\begin{document}

\preprint{FTUV--04--1029}

\title{The massive analytic invariant charge in QCD}

\author{A.V.~Nesterenko}
\email{nesterav@ific.uv.es}
\affiliation{\Valencia}
\affiliation{\Dubna}

\author{J.~Papavassiliou}
\email{Joannis.Papavassiliou@uv.es}
\affiliation{\Valencia}

\date{29 October 2004}

\begin{abstract}
The low energy behavior of a recently proposed model for the massive
analytic running coupling of QCD is studied. This running coupling
has no unphysical singularities, and in the absence of masses
displays infrared enhancement. The inclusion of the effects due to
the mass of the lightest hadron is accomplished by employing the
dispersion relation for the Adler $D$~function.  The presence of the
nonvanishing pion mass tames the aforementioned enhancement, giving
rise to a finite value for the running coupling at the origin. In
addition, the effective charge acquires a ``plateau-like'' behavior
in the low energy region of the timelike domain. This plateau is
found to be in agreement with a number of phenomenological models for
the strong running coupling.  The developed invariant charge is
applied in the processing of experimental data on the inclusive
$\tau$~lepton decay. The effects due to the pion mass play an
essential role here as well, affecting the value of the QCD scale
parameter~$\Lambda$ extracted from these data. Finally, the massive
analytic running coupling is compared with the effective coupling
arising from the study of Schwinger--Dyson equations, whose infrared
finiteness is due to a dynamically generated gluon mass. A
qualitative picture of the possible impact of the former coupling on
the chiral symmetry breaking is presented.
\end{abstract}


\maketitle

\section{Introduction}

     The theoretical analysis of strong interaction processes at low
energies represents a long-standing challenge for Quantum
Chromodynamics~(QCD). Whereas the discovery of asymptotic
freedom~\cite{AF} was followed by the rapid development of
perturbative tools for the detailed study of the ultraviolet region,
a reliable method for description of hadron dynamics in the infrared
domain is still missing. Given that many important QCD phenomena,
such as hadronization, quark confinement, chiral symmetry breaking,
and dynamical mass generation, are infrared in origin, one resorts to
the variety of models, in an attempt to obtain a consistent
quantitative description of the low energy dynamics.

     The renormalization group (RG) method~\cite{RG,BgSh} plays a
fundamental role in the framework of Quantum Field Theory~(QFT) and
its applications. In the case of QCD, in order to describe the
physics in the asymptotical ultraviolet region, one basically applies
the RG method together with perturbative calculations. In this case,
owing to the asymptotic freedom, \textit{a priori} unknown RG
functions can be parameterized by power series in the strong running
coupling.  Eventually, this leads to approximate solutions of the RG
equations, which are used in the quantitative analysis of the
high-energy processes. However, such solutions possess unphysical
singularities in the infrared domain, contradicting the general
principles of the local QFT, and complicating the theoretical
description and interpretation of the intermediate- and low-energy
experimental data.  Nonetheless, these difficulties, being artifacts
of the perturbative treatment of the RG method, can be circumvented
by judiciously incorporating nonperturbative information about the
hadron dynamics at low energies.

     It is worth mentioning several well--known examples of such
``synthesis''. The short--range part of the static quark--antiquark
potential can be calculated perturbatively~\cite{Peter}, while its
linear confining behavior at large distances is corroborated by both
the lattice results (see recent papers~\cite{LatQCD}) and the string
hadron models (see, e.g., book~\cite{Nest} and references therein).
These two inputs complement each other and form the so-called
``V--scheme''~\cite{VScheme} for the QCD effective charge, which has
proved to be successful in describing hadrons as bound states of
quarks~\cite{HadrSpectr}. The so-called ``I--scheme''~\cite{IScheme}
is constructed along the same lines. Here, the perturbative results
are supplied with the large distance behavior of the running
coupling, coming from the lattice study~\cite{UKQCD} of the
topological structure of the QCD vacuum. Interestingly enough, both
aforementioned schemes, although being based on different
assumptions, predict a similar infrared behavior for the strong
running coupling.  Furthermore, the latter also agrees with that of
the model for the QCD analytic invariant charge developed in
Refs.~\cite{PRD1,PRD2} (see also Refs.~\cite{Schrempp,Review} for the
details). There is also a number of methods which proceed from the
general properties of the perturbative power series for the QCD
observables in the framework of the renormalization group formalism.
For example, these are the ``optimized perturbation
theory''~\cite{OPT,OPTEpEm}, the method of effective
charges~\cite{EffChar}, the Brodsky-Lepage-Mackenzie convergence
criterion~\cite{BLM}, the ``optimal conformal mapping''
method~\cite{Fischer}, and the RG improvement of perturbative
calculations~\cite{Elias1,Elias2}.

     Another important source of nonperturbative information is
provided by the relevant dispersion relations. The latter, being
based on the ``first principles'' of the theory, supply one with the
definite analytic properties with respect to a given kinematic
variable of a physical quantity in hand. The idea of employing this
information together with perturbative treatment of the
renormalization group method forms the underlying concept of the
so--called ``analytic approach'' to~QFT. It was first proposed in the
framework of Quantum Electrodynamics~(QED) and applied to the study
of the invariant charge of the theory~\cite{AQED}. Here the principle
of causality implies the \KL spectral representation for the QED
running coupling. Hence, the latter has to be an analytic function in
the complex $q^2$-plane with the only cut along the
negative\footnote{A metric with signature $(-1, 1, 1, 1)$ is used, so
that positive $q^2$ corresponds to a spacelike momentum transfer.}
semiaxis of real~$q^2$. A number of authors (see, e.g.,
Ref.~\cite{Bjorken}) have argued that a similar method can also be
useful for studying non-Abelian theories.  Eventually, proceeding
from these motivations, the ``dispersive approach''~\cite{DMW} and
the ``analytic approach''~\cite{ShSol} to QCD have been developed.
According to the former one, the nonperturbative effects of the
strong interaction can be reliably captured at an inclusive level by
means of a quantity, which constitutes the effective extension of the
perturbative running coupling to the low energy scales. The analytic
approach to QCD has been successfully applied to the study of the
strong running coupling~\cite{ShSol,PRD2}, perturbative series for
the QCD observables~\cite{APTRev}, and some intrinsically
nonperturbative aspects of the strong
interaction~\cite{PRD1,Review,QCD03}. Some of the main advantages of
the latter approach are the absence of unphysical singularities and a
fairly good higher-loop and scheme stability of the outcoming
results.  Besides, in the framework of the analytic approach the
continuation of the ``spacelike'' theoretical predictions for the QCD
observables into the timelike domain, that is crucial for handling
the relevant experimental data, can be carried out in a
self-consistent way~\cite{MS97}.

     In general, the effects due to the masses of light hadrons (such
as $\pi$~meson) can be safely neglected only when one studies the
strong interaction processes at large momenta transferred. For
example, in order to relate the perturbative results with the high
energy experimental data on the electron--positron annihilation into
hadrons, the massless approximation of the dispersion relation for
the \ADf{} may be used (see Section~\ref{Sect:RCSLTL} for the
details).  But for the hadron dynamics in the infrared domain the
mass effects become substantial. Apparently, this is important for
the description of the low energy experimental data on the inclusive
$\tau$~lepton decay. Both, the results of perturbation theory and the
dispersion relation for the \ADf{} with the nonvanishing pion mass,
are vital here for properly processing these data.  However, no such
mass effects have been taken into account within the analytic
approach to QCD so far.

     The primary objective of this paper is to include the effects
due to the pion mass into the analytic approach to QCD.  The
incorporation of such mass effects is studied on the example of the
model for the analytic running coupling developed in
Refs.~\cite{PRD1,PRD2}. Therein, the imposition of the analyticity
requirement has eventually resulted in the infrared enhancement
(i.e., the singular behavior at~$q^2=0$) of the invariant charge in
hand.  In general, one might anticipate that the presence of masses
affects the low energy behavior of the strong running coupling.
Indeed, as we shall see, the aforementioned singularity is tamed down
by the pion mass, thus giving rise to a finite infrared limiting
value for the QCD effective charge. Apparently, it is important to
apply the developed model to the description of those sets of
experimental data, which display a particular sensitivity to the
infrared behavior of the QCD running coupling. It is also of
significant interest to examine, even at a qualitative level, the
applicability of the obtained invariant charge to the study of the
chiral symmetry breaking through \SD equations.

     The layout of the paper is as follows.
Section~\ref{Sect:RCSLTL} is devoted to the description of the strong
interaction processes in spacelike and timelike domains. This
material sets up the stage for the subsequent analysis of the
massless and massive cases. In Section~\ref{Sect:AQCD} the analytic
approach to QCD is overviewed, with a particular emphasis on the
massless model for the invariant charge of~\cite{PRD1,PRD2}. The
effects due to the pion mass are incorporated into the latter
approach in Section~\ref{Sect:MAQCD}. The basic features of the
massive strong running coupling in spacelike and timelike regions are
also studied therein. In Section~\ref{Sect:Tau} the developed model
for the invariant charge is applied to processing the experimental
data on the inclusive $\tau$~lepton decay, a reasonable estimation of
the QCD scale parameter $\Lambda$ being obtained. In
Section~\ref{Sect:CSB} the derived massive analytic charge is
compared with the effective charge arising from the study of the \SD
equations, whose infrared finiteness is due to a dynamically
generated gluon mass~\cite{Cornwall:1982zr}. A qualitative picture of
the possible impact of the former charge on the chiral symmetry
breaking is presented.  In Conclusions (Section~\ref{Sect:Concl}) the
basic results are summarized and further studies within this approach
are outlined.

\section{Strong running coupling in spacelike and timelike regions}
\label{Sect:RCSLTL}

     The consistent description of hadron dynamics in timelike
(Minkowskian) and spacelike (Euclidean) regions remains the subject
of intense studies. The strong interaction processes involving the
large spacelike momentum transfer $q^2>0$ (for instance, the deep
inelastic lepton--hadron scattering) can be examined perturbatively
in the framework of the RG method (see, e.g., Ref.~\cite{DISRG}).
However, in order to handle the processes which depend on the
timelike kinematic variable $s=-q^2>0$ (for example, hadronic width
of the $\tau$~lepton decay or total cross--section of the
electron--positron annihilation into hadrons), one first has to
relate the results of perturbation theory with the measured
quantities.  Obviously, the question what is the expansion parameter
for the QCD timelike processes arises at this stage~\cite{PenRos}.

     An indispensable method for the analysis of the strong
interaction processes in the timelike domain has been proposed by
Adler~\cite{Adler}, and further developed in
Refs.~\cite{Rad82,KrPi82}. In particular, it was argued that the
logarithmic derivative of the hadronic vacuum polarization
function~$\Pi(q^2)$
\begin{equation}
\label{AdlerDef}
D(q^2) = \frac{d\, \Pi(q^2)}{d \ln q^2},
\end{equation}
which is also known as the \ADf, provides a firm ground for comparing
the perturbative results with the experimental data on the $\epem$
annihilation into hadrons. Specifically, the dispersion
relation~\cite{Adler}
\begin{equation}
\label{AdlerDisp}
D(q^2) = q^2 \int_{4\mpi^2}^{\infty}
\frac{R(s)}{(s + q^2)^2}\, d s
\end{equation}
embodies the required link between the measurable ratio of two
cross--sections~\cite{Repem}
\begin{eqnarray}
R(s) &=& \frac{\sigma\left(\epem \to \text{hadrons}; s\right)}
{\sigma\left(\epem \to \mu^{+}\mu^{-}; s\right)}
\nonumber \\*
&=& \frac{1}{2 \pi i} \lim_{\varepsilon \to 0_{+}}
\left[\Pi(- s + i \varepsilon) - \Pi(- s - i \varepsilon)\right]
\label{RepemDef}
\end{eqnarray}
and the \ADf, which can be calculated perturbatively.  In
Eq.~(\ref{RepemDef}) $s$~denotes the center-of-mass energy of the
annihilation process. Thus, one can continue the perturbative results
for~$D(q^2)$ into the timelike domain by making use of the relation
inverse to Eq.~(\ref{AdlerDisp})
\begin{equation}
\label{AdlerInv}
R(s) = \frac{1}{2 \pi i} \lim_{\varepsilon \to 0_{+}}
\int_{s + i \varepsilon}^{s - i \varepsilon}
D(-\zeta) \, \frac{d \zeta}{\zeta},
\end{equation}
where the integration path lies in the region of analyticity of the
function~$D(-\zeta)$, see also Refs.~\cite{Rad82,GKL91}.

     So far, there is no systematic method for calculating the \ADf.
Nevertheless, its asymptotic ultraviolet behavior at $q^2 \to \infty$
can be computed perturbatively. There, the effects due to the masses
of light hadrons can be neglected, and the \ADf{} of
Eq.~(\ref{AdlerDef}) is usually approximated by the power series in
the strong running coupling~$\al{}{s}(q^2)$
\begin{equation}
\label{AdlerGen}
D(q^2) = \Nc \sum_{f} Q_f^2 \left[1 + d(q^2) \right],
\end{equation}
where $\Nc = 3$ is the number of colors, $Q_f$ stands for the charge
of the $f$-th quark,
\begin{equation}
\label{AdlerPert}
d(q^2) \simeq
d_1\left[\frac{\al{}{s}(q^2)}{\pi}\right] +
d_2\left[\frac{\al{}{s}(q^2)}{\pi}\right]^2 + \ldots,
\end{equation}
$d_1 = 1$, $\,d_2 \simeq 1.9857 - 0.1153 \nf$, and $\nf$ is the
number of active quarks, see Refs.~\cite{GKL91,SurSa91} for the
details.

     Thus, in order to compare the perturbative results with the
timelike experimental data, one first has to perform on
Eq.~(\ref{AdlerPert}) the integral transformation given in
Eq.~(\ref{AdlerInv}). It is worthwhile to underscore that this
procedure distorts the perturbative power series for the \ADf{}
drastically, since both real and imaginary parts of the running
coupling $\al{}{s}(q^2)$ contribute to Eq.~(\ref{AdlerInv}).
Ultimately, the continuation presented in Eq.~(\ref{AdlerInv})
results in a ``non-power'' expansion for $R(s)$, and even in the deep
ultraviolet asymptotic $|q^2| \to \infty$ the functions $D(q^2)$ and
$R(s)$ are different, starting from the three-loop level, due to the
so-called $\pi^2$--terms. Nonetheless, the ``naive'' extrapolation of
the strong running coupling to the timelike domain $\hal{}{}(s) =
\al{}{s}(|q^2|)$ is also allowed for the perturbative expansion of
Eq.~(\ref{AdlerPert}), but only if one restricts oneself to the deep
ultraviolet limit $|q^2|\to\infty$ of the one- or two-loop levels
(see Refs.~\cite{Bjorken,APTRev,MS97,Rad82,KrPi82,DV01,APTTau} for
the details).

     Since the integral transformation~(\ref{AdlerInv}) of the
perturbative results has to be carried out every time one deals with
the timelike strong interaction processes, for practical purposes it
is convenient to define~\cite{MS97} the timelike effective charge
$\hal{}{}(s)$ in the same way, as $R(s)$ relates with~$D(q^2)$:
\begin{equation}
\label{TLviaSL}
\hal{}{}(s) = \frac{1}{2 \pi i} \lim_{\varepsilon \to 0_{+}}
\int_{s + i \varepsilon}^{s - i \varepsilon}
\al{}{}(- \zeta) \, \frac{d \zeta}{\zeta}.
\end{equation}
In what follows the strong running coupling in the spacelike domain
is denoted by $\al{}{}(q^2)$, and in the timelike domain by
$\hal{}{}(s)$.  Obviously, the inverse relation between these
effective charges\footnote{The case of the massless pion $\mpi=0$ was
studied in Refs.~\cite{MS97,DV01,APTTau}.}
\begin{equation}
\label{SLviaTL}
\alpha(q^2) = q^2 \int_{4 \mpi^2}^{\infty}
\frac{\hal{}{}(s)}{(s + q^2)^2}\, d s
\end{equation}
holds as well\footnote{The relations~(\ref{TLviaSL})
and~(\ref{SLviaTL}) are not valid for the perturbative running
coupling $\al{}{s}(q^2)$ because of the unphysical singularities of
the latter, see Section~\ref{Sect:AQCD} for the details.}.  It is
important to emphasize that for a detailed description of the
infrared hadron dynamics the pion mass cannot be neglected in
Eqs.~(\ref{AdlerDisp}) and~(\ref{SLviaTL}).

     Apparently, for the self--consistency of the method described
above, one first has to bring the perturbative approximation for the
\ADf{} in Eq.~(\ref{AdlerPert}) to conformity with the dispersion
relation of Eq.~(\ref{AdlerDisp}). This is of a great significance
when one intends to study the QCD experimental data in the
intermediate- and low-energy regions. Indeed, the integral
representation in Eq.~(\ref{AdlerDisp}) implies the definite analytic
properties in~$q^2$ variable for the \ADf. For example, in the
massless limit ($\mpi=0$), it has to be an analytic function in the
complex $q^2$-plane with the only cut $- \infty < q^2 \le 0$ along
the negative semiaxis of real~$q^2$. However, the approximation of
the right hand-side of Eq.~(\ref{AdlerGen}) by the perturbative
expansion in the strong running coupling given in
Eq.~(\ref{AdlerPert}) obviously violates this condition.
Nevertheless, this discrepancy can be eliminated in the framework of
the analytic approach to QCD, which is discussed in the next section.

\section{Massless analytic running coupling}
\label{Sect:AQCD}

     As has already been mentioned in the Introduction, the
dispersion relations play a central role in the description of hadron
dynamics. Indeed, the general principles of the local QFT (such as
causality, spectrality, unitarity) are captured by the relevant
integral representations. These are, for instance, the dispersion
relation for the \ADf~(\ref{AdlerDisp}) and the \JLD
representation~\cite{JLD} for the structure function of the deep
inelastic lepton--hadron scattering processes.  In turn, the
dispersion relations provide one with a certain nonperturbative
information about the quantity in hand, in particular, with the
definite analytic properties in the kinematic variable. Undoubtedly,
the latter should be taken into account when one intends to venture
beyond the realm of perturbation theory.

     It has recently been argued~\cite{DMW,ShSol} that for the QCD
invariant charge $\alpha(q^2)$ the \KL spectral representation
\begin{equation}
\label{KLAn}
\alpha(q^2) = \int_{0}^{\infty}
\frac{\varrho(\sigma)}{\sigma + q^2} \, d \sigma
\end{equation}
must hold in the absence of masses. The condition~(\ref{KLAn}) is
identical to that needed\footnote{In the limit of the massless
pion~$\mpi=0$.} for bringing the perturbative approximation of the
\ADf{} in Eq.~(\ref{AdlerPert}) to conformity with its dispersion
relation~(\ref{AdlerDisp}), also enforcing the validity of
Eqs.~(\ref{TLviaSL}) and~(\ref{SLviaTL}).  However, there are several
ways to incorporate the analyticity requirement of Eq.~(\ref{KLAn})
for the QCD running coupling into the RG formalism.  In other words,
the perturbative asymptotic behavior of $\al{}{s}(q^2)$ when
$q^2\to\infty$, together with the integral
representation~(\ref{KLAn}), is not enough to uniquely determine the
relevant spectral density~$\varrho(\sigma)$.  Eventually, this
ambiguity has given rise to different models for the strong running
coupling within the analytic approach to QCD (discussion of this
issue can also be found in Refs.~\cite{PRD2,Review,MPLA2,DV02,DV04}).

     This section is devoted to a brief overview of one of the
massless models for the QCD analytic invariant
charge~\cite{PRD1,PRD2}. This model shares all the advantages of the
analytic approach, namely, it contains no unphysical singularities,
and displays good higher loop convergence and mild dependence on the
subtraction scheme.  Besides, the running coupling of
Refs.~\cite{PRD1,PRD2} was successful in the description of a wide
range of QCD phenomena \cite{Review,QCD03}.  Furthermore, it is of a
particular interest to note that this model has recently been
re-derived, proceeding from completely different
motivations~\cite{Schrempp}.

     In the framework of perturbation theory the RG equation for the
QCD invariant charge $\alpha(\mu^2)= g^2(\mu^2)/(4\pi)$ at the
$\ell$-loop level takes the form
\begin{equation}
\label{RGEqPert}
\frac{d\,\ln \al{(\ell)}{s}(\mu^2)}{d\,\ln \mu^2} = -
\sum_{j=0}^{\ell-1} \beta_{j}
\left[\frac{\al{(\ell)}{s}(\mu^2)}{4 \pi}\right]^{j+1}.
\end{equation}
Here $\al{(\ell)}{s}(\mu^2)$ denotes the $\ell$-loop perturbative
running coupling, $\beta_{j}$ stands for the $\beta$~function
expansion coefficient ($\beta_{0} = 11 - 2 \nf / 3,\,$ $\beta_{1}=102
- 38 \nf / 3, \text{...}\,$), and $\nf$ is the number of active
quarks. It is well-known that the solutions to Eq.~(\ref{RGEqPert})
have unphysical singularities in the infrared domain at any loop
level. Specifically, the Landau pole appears at the one-loop level,
whereas the higher loop corrections introduce additional
singularities of the cut type into expression for the QCD invariant
charge. In turn, this contradicts the fundamental principles of the
local QFT, violating the representation given in Eq.~(\ref{KLAn}).

     In order to resolve this difficulty, in the framework of the
developed model~\cite{PRD1,PRD2} the analyticity requirement was
imposed on the $\beta$~function perturbative
expansion\footnote{Unlike the Shirkov--Solovtsov model~\cite{ShSol},
where the analyticity requirement~(\ref{DefAn}) was imposed on the
perturbative running coupling~$\al{}{s}(q^2)$ itself. In turn, this
has led to a spectral density somewhat different from that of
Eq.~(\ref{SpDnsHL}), and, consequently, to different properties of
the QCD effective charge in the infrared domain, see, e.g.,
Refs.~\cite{PRD2,Review,DV02,DV04} for the details.}
\begin{equation}
\label{RGEqAn}
\frac{d\,\ln \al{(\ell)}{an}(\mu^2)}{d\,\ln \mu^2} =
- \left\{\sum_{j=0}^{\ell-1} \beta_{j}
\left[\frac{\al{(\ell)}{s}(\mu^2)}{4 \pi}
\right]^{j+1}\right\}\ind{}{$\!$an}.
\end{equation}
In this equation $\al{(\ell)}{an}(\mu^2)$ is the $\ell$-loop analytic
invariant charge and the braces $\bigl\{\,\ldots\,\bigr\}\ind{}{an}$
denote the ``analytization'' of the expression contained in
them~\cite{ShSol}:
\begin{eqnarray}
\Bigl\{A(q^2)\Bigr\}\ind{}{$\!$an} & = &
\frac{1}{2\pi i}\int_{0}^{\infty}
\lim_{\varepsilon \to 0_{+}} \Bigl[A(-\sigma - i \varepsilon)
\nonumber \\*
&& - A(-\sigma + i \varepsilon)\Bigr]
\frac{d \sigma}{\sigma + q^2}.
\label{DefAn}
\end{eqnarray}
It is worth noting here that the way of incorporating the analyticity
requirement into the RG method given in Eq.~(\ref{RGEqAn}) is
consistent with the general definition of the QCD invariant charge,
see Refs.~\cite{Review,MPLA2}.

     At the one-loop level the RG equation~(\ref{RGEqAn}) for the
analytic invariant charge can be solved explicitly~\cite{PRD1}:
\begin{equation}
\label{AIC1L}
\al{(1)}{an}(q^2) = \fpb\, \frac{z - 1}{z \, \ln z},
\qquad z = \frac{q^2}{\Lambda^2}.
\end{equation}
At the higher loop levels only the integral representation for the
analytic running coupling has been derived. So, at the $\ell$-loop
level the solution to Eq.~(\ref{RGEqAn}) acquires the
form~\cite{PRD2,MPLA2}:
\begin{equation}
\label{AICHL}
\al{(\ell)}{an}(q^2) = \fpb\,\frac{z-1}{z \, \ln z}\,
\exp\!\left[\int_{0}^{\infty}\! {\cal P}^{(\ell)}(\sigma)\,
\ln\!\left(1 + \frac{\sigma}{z}\right) \frac{d \sigma}{\sigma}\right],
\end{equation}
where ${\cal P}^{(\ell)}(\sigma) = \Ro{\ell}{\sigma} -
\Ro{1}{\sigma}$ and
\begin{eqnarray}
\Ro{\ell}{\sigma} &=& \frac{1}{2 \pi i}\,
\lim_{\varepsilon \to 0_{+}} \sum_{j=0}^{\ell-1}
\frac{\beta_{j}}{(4 \pi)^{j+1}}\Bigl\{
\left[\al{(\ell)}{s}(-\sigma - i \varepsilon)\right]^{j+1}
\nonumber \\*
&& - \left[\al{(\ell)}{s}(-\sigma + i \varepsilon)\right]^{j+1}
\Bigr\}.
\end{eqnarray}

     The obtained massless running coupling~(\ref{AICHL}) has the
correct analytic properties in the $q^2$~variable demanded in
Eq.~(\ref{KLAn}), namely, it has the only cut $q^2 \le 0$ along the
negative semiaxis of real~$q^2$. In particular, the latter follows
from the \KL integral representation that holds for the invariant
charge~(\ref{AICHL}):
\begin{equation}
\label{AICHLKL}
\al{(\ell)}{an}(q^2) = \fpb \int_{0}^{\infty}
\frac{\ro{\ell}(\sigma)}{\sigma + z}\, d \sigma.
\end{equation}
In this equation $\ro{\ell}(\sigma)$ denotes the $\ell$-loop spectral
density
\begin{eqnarray}
\ro{\ell}(\sigma) &=& \ro{1}(\sigma) \,
\exp\!\!\left[\int_{0}^{\infty} {\cal P}^{(\ell)}(\zeta) \,
\ln \left| 1 - \frac{\zeta}{\sigma}\right| \,
\frac{d \zeta}{\zeta} \right] \nonumber \nopagebreak \\
&& \times \left[\cos \psi^{(\ell)}(\sigma) +
\frac{\ln \sigma}{\pi} \sin \psi^{(\ell)}(\sigma) \right],
\label{SpDnsHL}
\end{eqnarray}
where
\begin{equation}
\psi^{(\ell)}(\sigma) = \pi \int_{\sigma}^{\infty}\!
{\cal P}^{(\ell)}(\zeta) \, \frac{d \zeta}{\zeta},
\end{equation}
and
\begin{equation}
\label{SpDns1L}
\ro{1}(\sigma) =
\left(1+\frac{1}{\sigma}\right)\frac{1}{\ln^2\!\sigma+\pi^2}
\end{equation}
is the one-loop spectral density. In the exponent of
Eq.~(\ref{SpDnsHL}) the principal value of the integral is assumed
(see Refs.~\cite{PRD2,Review} for the details).

     The massless analytic running coupling of Eq.~(\ref{AICHL})
possesses a number of appealing features. First of all, it has no
unphysical singularities at any loop level, and contains no
adjustable parameters\footnote{It is worth noting here that the
Shirkov--Solovtsov running coupling~\cite{ShSol} has no adjustable
parameters, either. So, both these models are the ``minimal'' ones in
this sense.}. Thus, similarly to the perturbative approach, the QCD
scale parameter $\Lambda$ remains the basic characterizing quantity
of the theory. In addition, the invariant charge~(\ref{AICHL})
incorporates the ultraviolet asymptotic freedom with the infrared
enhancement in a single expression, which plays an essential role in
applications of the developed model to the description of the
quenched lattice simulation data~\cite{Schrempp,QCD03,Vr}. Moreover,
this analytic running coupling has universal asymptotics both in the
ultraviolet and infrared regions at any loop level, and displays a
good higher loop and scheme stability. The detailed analysis of the
properties of the invariant charge~(\ref{AICHL}) and its applications
can be found in Refs.~\cite{Review,QCD03,MPLA2,MPLA1}.

\begin{figure}[t]
\includegraphics[width=85mm]{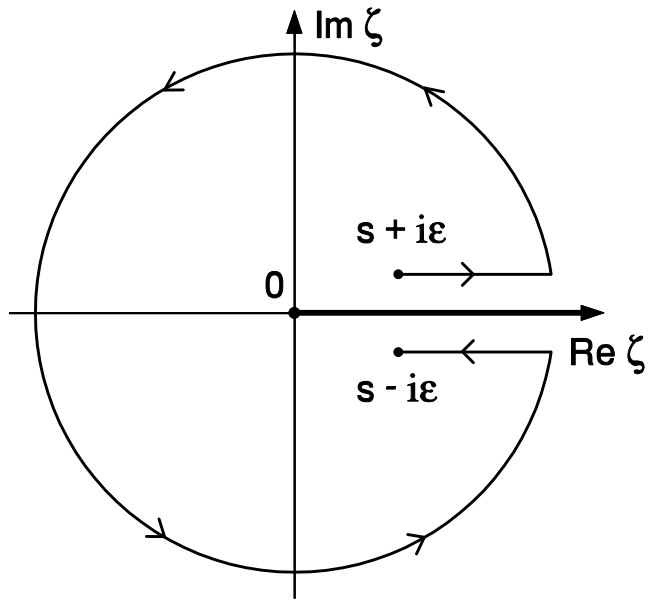}
\caption{The integration contour in Eq.~(\protect\ref{TLviaSL}) for
the massless case. The physical cut of the strong running coupling
$\alpha(-\zeta)$ (see Eq.~(\ref{KLAn})) is shown along the positive
semiaxis of real~$\zeta$.}
\label{Plot:ContTL}
\end{figure}

     As has been discussed in Section~\ref{Sect:RCSLTL}, for the
consistent description of a number of strong interaction processes
one has to employ the continuation of the QCD effective charge to the
timelike region, in the way given in Eq.~(\ref{TLviaSL}).  For the
massless case under consideration it is convenient to choose the
integration contour in Eq.~(\ref{TLviaSL}) in the form presented in
Figure~\ref{Plot:ContTL}. Eventually, this leads to the following
extension of the invariant charge of Eq.~(\ref{AICHL}) to the
timelike domain~\cite{PRD2}
\begin{equation}
\label{AICTL}
\hal{(\ell)}{an}(s) = \fpb\, \int_{w}^{\infty}\!
\ro{\ell}(\sigma)\,\frac{d \sigma}{\sigma},
\qquad w = \frac{s}{\Lambda^2},
\end{equation}
where $s=-q^2>0$, and the spectral density $\ro{\ell}(\sigma)$ is
defined in Eq.~(\ref{SpDnsHL}). The obtained result supports the
hypothesis due to Schwinger~\cite{Schwinger,Milton} concerning the
proportionality between the $\beta$~function and the relevant
spectral density (see also Ref.~\cite{MS97}).

     The one-loop effective charge of Eq.~(\ref{AICTL}) has the
following asymptotic in the high energy limit $s \to \infty$:
\begin{equation}
\label{AICTLUV}
\hal{(1)}{an}(s) \simeq \fpb \, \frac{1}{\ln w} \,
\left[1 - \frac{\pi^2}{3} \frac{1}{\ln^2\!w} +
\cor\left(\frac{1}{\ln^4\!w}, \frac{1}{w}\right)\right].
\end{equation}
On the one hand, this running coupling has the correct ultraviolet
behavior, determined by the asymptotic freedom. On the other hand,
the so-called $\pi^2$--terms have also appeared in the
expansion~(\ref{AICTLUV}). As it was noticed in
Section~\ref{Sect:RCSLTL}, these terms play a key role in the
description of the strong interaction processes in the timelike
domain. It is interesting to note that, similarly to the
``spacelike'' running coupling in the massless case of
Eq.~(\ref{AIC1L}), the one-loop effective charge~(\ref{AICTL}) also
has an enhancement in the infrared domain (see
Refs.~\cite{PRD2,Review}):
\begin{equation}
\label{AICTLIR}
\hal{(1)}{an}(s) \simeq \fpb \,\frac{1}{w\,\ln^2\!w},
\qquad s \to 0.
\end{equation}
However, the type of this singularity differs from that of the
invariant charge in Eq.~(\ref{AIC1L}) by the logarithmic factor.
Nevertheless, it is precisely this feature of the timelike running
coupling that enables one to handle the integrals of a specific form
over the infrared region, and in particular, to process the
experimental data on the inclusive semileptonic branching ratio for
the case of the massless pion.  In turn, the latter provides one with
the relevant estimation of the QCD scale parameter $\Lambda = (508
\pm 61)$~MeV, see Section~\ref{Sect:Tau} for the details.

\begin{figure}[t]
\includegraphics[width=85mm]{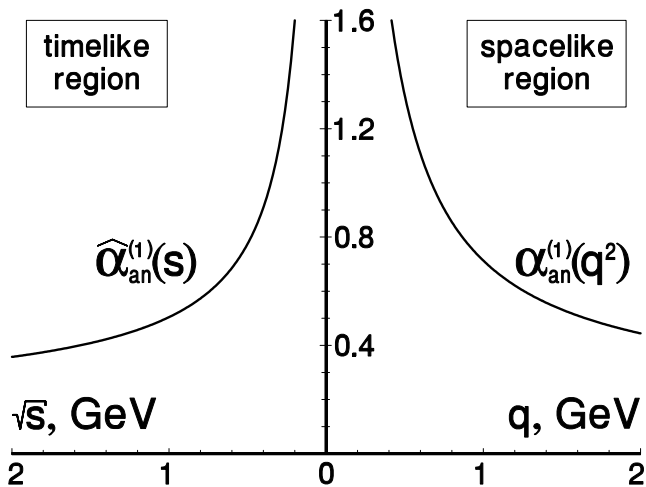}
\caption{The one-loop massless analytic running coupling in the
spacelike (Eq.~(\protect\ref{AICHLKL}), $q^2>0$) and timelike
(Eq.~(\protect\ref{AICTL}), $s=-q^2>0$) regions. The values
of parameters are: $\Lambda = 508$~MeV, $\nf=2$ active quarks.}
\label{Plot:AICTL}
\end{figure}

     The plots of the functions $\al{(1)}{an}(q^2)$ and
$\hal{(1)}{an}(s)$ are shown in Figure~\ref{Plot:AICTL}. In the
ultraviolet limit these expressions have identical behavior
determined by the asymptotic freedom. However, there is an asymmetry
between them in the intermediate and low energy regions. Thus, the
relative difference between these effective charges is about several
percent at the scale of the Z~boson mass, and increases when
approaching the infrared domain. Evidently, this circumstance has to
be taken into account when one handles the experimental data (see
also review~\cite{Review} and references therein for the details).

     It is worthwhile to emphasize that the mass effects have not
been included in the formulation of the models for the strong running
coupling in the framework of the analytic approach to QCD, so far.
Thus, the obtained results can be applied, for example, to the study
of the experimental data at high energies, where the masses of the
lightest hadrons can be neglected, the pure gluodynamics, and the
quenched lattice simulation data (see also
Refs.~\cite{Schrempp,Review}). However, for the detailed description
of the infrared hadron dynamics, the mass effects have to be
incorporated into the analytic approach to QCD.  The next section is
devoted to this task.

\section{Massive analytic effective charge}
\label{Sect:MAQCD}

     As has been noticed in previous sections, the $\pi$~meson plays
a crucial part in the description of the strong interaction processes
at low energies.  So far the main thrust of the analytic approach to
QCD has focused on eliminating intrinsic difficulties of perturbation
theory, such as the unphysical singularities of the strong running
coupling (see Section~\ref{Sect:AQCD}).  On the other hand, mass
effects within this formalism remain largely unexplored, thus far.
Therefore, the objective of this section is to incorporate the
effects due to the pion mass into the analytic approach to QCD.

     Evidently, the original dispersion relation for the
\ADf~\cite{Adler} (see Eq.~(\ref{AdlerDisp})) with the nonvanishing
mass of the $\pi$~meson is the proper object to study here. Indeed,
Eq.~(\ref{AdlerDisp}) implies definite analytic properties in the
$q^2$ variable for $D(q^2)$. Namely, the latter has to be an analytic
function in the complex $q^2$-plane with the only cut beginning at
the two--pion threshold $-\infty < q^2 \le -4\mpi^2$ along the
negative semiaxis of real~$q^2$. However, its approximation in
Eq.~(\ref{AdlerPert}) violates this condition due to the spurious
singularities of the perturbative running coupling~$\al{}{s}(q^2)$.
Nevertheless, this disagreement can be avoided by imposing the
analyticity requirement of the form\footnote{The spectral function
$\varkappa(\sigma)$ in Eq.~(\ref{AdlerPertAnM}) is supposed to
capture the known perturbative contributions to~$d(q^2,\mpi^2)$.}
\begin{equation}
\label{AdlerPertAnM}
d(q^2, \mpi^2) = \int_{4 \mpi^2}^{\infty}
\frac{\varkappa(\sigma)}{\sigma + q^2}\, d \sigma
\end{equation}
on the right hand-side of Eq.~(\ref{AdlerPert}). Therefore, the QCD
effective charge itself has to satisfy the integral representation
\begin{equation}
\label{KLAnM}
\alpha(q^2, \mpi^2) = \int_{4 \mpi^2}^{\infty}
\frac{\varrho(\sigma)}{\sigma + q^2}\, d \sigma
\end{equation}
in this case as well. Otherwise, one would encounter a contradiction
between the dispersion relation for the \ADf{} of
Eq.~(\ref{AdlerDisp}) and its approximation given in
Eq.~(\ref{AdlerPert}).  Besides, the condition~(\ref{KLAnM}) enforces
the validity of relations~(\ref{TLviaSL}) and~(\ref{SLviaTL}) for the
case of the nonvanishing pion mass.

     In general, there are several models for the invariant charge
within the analytic approach to QCD (see Section~\ref{Sect:AQCD} for
the details).  This is so by virtue of the fact that the behavior of
the strong running coupling $\al{}{s}(q^2)$ at the ultraviolet
asymptotic, which is known from perturbation theory, together with
the analyticity requirement of the form of Eq.~(\ref{KLAn}) or
Eq.~(\ref{KLAnM}), is not enough to uniquely determine the relevant
spectral density $\varrho(\sigma)$. The model for the analytic
invariant charge~\cite{PRD1,PRD2} has proved to be successful in the
description of the strong interaction processes of both perturbative
and intrinsically nonperturbative nature~\cite{Review,QCD03}. We
shall therefore adopt the spectral density of Eq.~(\ref{SpDnsHL}) in
what follows.

     Thus, one arrives at the following integral representation for
the massive analytic invariant charge (see also
Refs.~\cite{NPQCD04,QCD04})
\begin{equation}
\label{AICMHL}
\al{(\ell)}{an}(q^2, \mpi^2) = \fpb\int_{\chi}^{\infty}
\frac{\ro{\ell}(\sigma)}{\sigma + z}\, d\sigma,
\qquad z=\frac{q^2}{\Lambda^2},
\end{equation}
where $\ro{\ell}(\sigma)$ denotes the $\ell$-loop spectral density of
Eq.~(\ref{SpDnsHL}) and $\chi=4\mpi^2/\Lambda^2$. It is worth noting
from the very beginning that the nonvanishing mass of the $\pi$~meson
drastically affects the low energy behavior of this strong running
coupling. Indeed, instead of the infrared enhancement in the massless
case of Eq.~(\ref{AICHL}), one has here the infrared finite limiting
value for the massive invariant charge in Eq.~(\ref{AICMHL}),
\begin{equation}
\label{AICMIR}
\al{(\ell)}{0} = \fpb\int_{\chi}^{\infty}
\ro{\ell}(\sigma) \frac{d\sigma}{\sigma},
\end{equation}
which depends on the value of the pion mass.  At the ultraviolet
asymptotic, where the nonperturbative contributions are negligible,
the result of Eq.~(\ref{AICMHL}) tends to the perturbative running
coupling~$\al{(\ell)}{s}(q^2)$:
\begin{equation}
\label{AICMUV}
\al{(\ell)}{an}(q^2, \mpi^2) \simeq \al{(\ell)}{s}(q^2) +
\cor\left[\frac{\Lambda^2}{q^2},
\frac{\Lambda^2}{q^2}\frac{1}{\ln(q^2/\Lambda^2)},
\frac{4\mpi^2}{q^2}\right].
\end{equation}
In this equation the limits $q^2 \gg \Lambda^2$ and $q^2 \gg 4\mpi^2$
are assumed. In particular, the one-loop effective charge of
Eq.~(\ref{AICMHL}) reads for $q^2 \gg 4\mpi^2$
\begin{eqnarray}
\al{(1)}{an}(q^2, \mpi^2) & \simeq & \al{(1)}{s}(q^2) -
\fpb \frac{1}{z \ln z}
\nonumber \\* & &
- \frac{4}{\bz} \frac{1}{z} \left[\frac{\pi}{2} +
(\chi+1)\arctan\left(\frac{\ln\chi}{\pi}\right)\right]
\nonumber \\* & &
+ \frac{4}{\bz} \frac{1}{z} \int_{-\infty}^{\ln\chi}
e^{y}\arctan\left(\frac{y}{\pi}\right)\, d y.
\end{eqnarray}
It is worthwhile to mention also that in the limit of massless pion
$\mpi=0$ the effective charge~(\ref{AICMHL}) coincides with the
running coupling of Eq.~(\ref{AICHL}).

\begin{figure}[t]
\includegraphics[width=85mm]{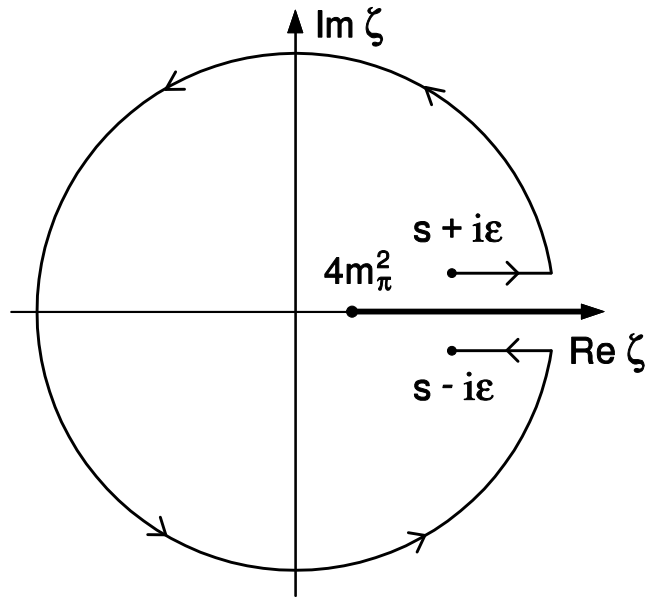}
\caption{The integration contour in Eq.~(\protect\ref{TLviaSL}) for
the case of the massive pion. The physical cut of the effective
charge $\alpha(-\zeta, \mpi^2)$ (see Eq.~(\ref{KLAnM})) is shown
along the positive semiaxis of real~$\zeta$.}
\label{Plot:ContTLM}
\end{figure}

     In order to handle the strong interaction processes involving
the timelike kinematic variable one first has to relate the
experimental data with the perturbative results (see
Section~\ref{Sect:RCSLTL}). For practical purposes it is convenient
to employ here the extension of the spacelike running coupling to the
timelike domain given by Eq.~(\ref{TLviaSL}).  The analytic
properties in the $q^2$~variable of the QCD invariant charge
$\alpha(q^2)$ are different for the massless~(\ref{KLAn}) and
massive~(\ref{KLAnM}) cases (see Figures~\ref{Plot:ContTL}
and~\ref{Plot:ContTLM}, respectively). Thus, the
continuation~(\ref{TLviaSL}) of the massive strong running
coupling~(\ref{AICMHL}) to the timelike region results in (see
Refs.~\cite{NPQCD04,QCD04} also)
\begin{equation}
\label{AICMTL}
\hal{(\ell)}{an}(s, \mpi^2) = \fpb\int_{w}^{\infty}
\theta(\sigma - \chi)\, \ro{\ell}(\sigma) \frac{d \sigma}{\sigma},
\qquad w=\frac{s}{\Lambda^2},
\end{equation}
where $s = -q^2 \ge 0$, $\,\theta(x)$ stands for the Heaviside
step-function (see, e.g., Ref.~\cite{AS}), $\ro{\ell}(\sigma)$ is the
$\ell$-loop spectral density defined in Eq.~(\ref{SpDnsHL}), and
$\chi=4\mpi^2/\Lambda^2$.

\begin{figure}[t]
\includegraphics[width=85mm]{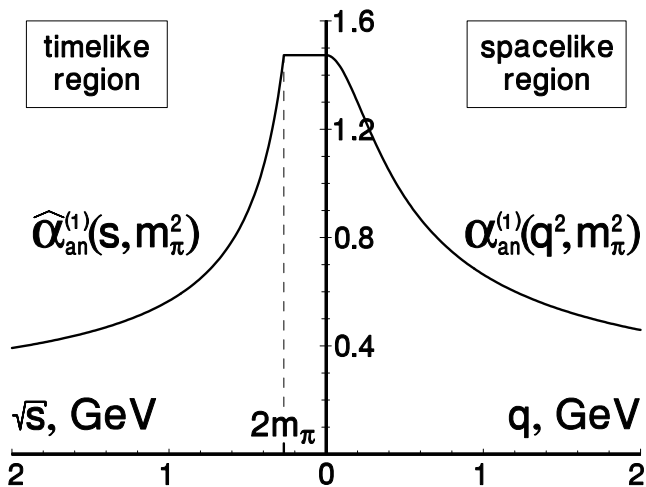}
\caption{The one--loop massive analytic effective charge in the
spacelike and timelike domains (Eqs.~(\protect\ref{AICMHL})
and~(\protect\ref{AICMTL}), respectively). The values of parameters
are: $\Lambda = 623$~MeV, $\nf=2$ active quarks.}
\label{Plot:AICM}
\end{figure}

     Let us address now the basic features of the running coupling in
Eq.~(\ref{AICMTL}). First of all, it is very interesting to note here
that the effective charges of Eq.~(\ref{AICMHL}) and
Eq.~(\ref{AICMTL}) have a common finite value in the infrared limit
$|q^2| \to 0$, given by Eq.~(\ref{AICMIR}). Second, the timelike
massive effective coupling of Eq.~(\ref{AICMTL}) has the
``plateau--like'' behavior in the deep infrared domain:
\begin{equation}
\label{Plateau}
\hal{(\ell)}{an}(s,\mpi^2) = \al{(\ell)}{0},
\qquad 0 \le \sqrt{s} \le 2 \mpi.
\end{equation}
Besides, for $\sqrt{s}>2\mpi$ there is no difference between the
massless and massive timelike running couplings of Eq.~(\ref{AICTL})
and Eq.~(\ref{AICMTL}), respectively, since the mass of the
$\pi$~meson affects the timelike effective charge~(\ref{AICMTL}) only
in the region $\sqrt{s}\le2\mpi$, where it does not run (see
Figure~\ref{Plot:AICM}). Therefore, in the ultraviolet asymptotic
$s\to\infty$, the expansion~(\ref{AICTLUV}), which accounts for the
$\pi^2$-terms, also holds for the running coupling of
Eq.~(\ref{AICMTL}). The hypothesis due to
Schwinger~\cite{Schwinger,Milton} concerning the proportionality
between the $\beta$~function and the relevant spectral density holds
for the massive timelike effective charge of Eq.~(\ref{AICMTL}) as
well.  Apparently, in the limit of vanishing pion mass $\mpi \to 0$
the results of this section reproduce the massless case described in
Section~\ref{Sect:AQCD}.

     It is worth noting here that some other models for the QCD
effective charge also display a plateau similar to~(\ref{Plateau}) in
the infrared domain. In particular, the aforementioned optimized
perturbation theory method~\cite{OPTEpEm} predicts the stagnation of
the timelike effective coupling in the region $\sqrt{s} \lesssim
300$~MeV, in striking coincidence with the result obtained in
Eq.~(\ref{Plateau}). Moreover, the so-called ``H--model'', with a
similar freezing of the effective charge to a constant value in the
infrared domain, has proved to be useful in studying of the dynamical
chiral symmetry breaking (see, e.g., Ref.~\cite{Hig84}).

\section{Inclusive $\tau$~lepton decay}
\label{Sect:Tau}

     In order to draw a quantitative conclusion on the low energy
behavior of a model for the strong running coupling, one needs the
relevant estimation of the QCD scale parameter~$\Lambda$. The latter
can be extracted, for example, from the experimental data on the
strong interaction processes. Among them, the measurement of the
$\tau$~decay width is most suitable for our purposes, since these
data are fairly precise, and this process probes the infrared hadron
dynamics at energies below the $\tau$~lepton mass $0 \le \sqrt{s} \le
\MTau$.  Let us turn now to the study of this hadron process,
restricting ourselves to the one--loop level at this stage.

     The experimentally measurable quantity here is the inclusive
semileptonic branching ratio
\begin{equation}
\label{RTauDef}
R_{\tau} =
\frac{\Gamma(\tau^{-} \to \text{hadrons}^{-}\, \nu_{\tau})}
{\Gamma(\tau^{-} \to e^{-}\, \bar\nu_{e}\, \nu_{\tau})}.
\end{equation}
One can split this ratio into three parts, namely
$R_{\tau}=\RTau{V}+\RTau{A}+\RTau{S}$. The terms $\RTau{V}$ and
$\RTau{A}$ account for the contributions to Eq.~(\ref{RTauDef}) of
the decay modes with the light quarks only, and they correspond to
the vector (V) and axial--vector~(A) quark currents, respectively.
The accuracy of the experimental measurement of these terms is
several times higher than that of the strange width ratio $\RTau{S}$,
which accounts for the contribution to Eq.~(\ref{RTauDef}) of the
decay modes with the $s$~quark. Thus, let us proceed with the
nonstrange part of the ratio $R_{\tau}$~(\ref{RTauDef}) associated
with the vector quark currents
\begin{equation}
\label{RTauVDef}
\RTau{V} = \frac{\Nc}{2}\,|\Vud|^2\, \Sew \left(1 +
\delta'_{\text{\tiny EW}} + \dQCD{} \right),
\end{equation}
see Refs.~\cite{BDP,BNP,APTTau} for detailed discussion of this
issue. In Eq.~(\ref{RTauVDef}) $\Nc=3$ is the number of colors,
$|\Vud|=0.9738 \pm 0.0005$ denotes the Cabibbo--Kobayashi--Maskawa
matrix element~\cite{PDG04}, $\Sew = 1.0194 \pm 0.0050$ and
$\delta'_{\text{\tiny EW}}=0.0010$ are the electroweak
corrections~\cite{BNP,EWF}, and $\dQCD{}$ stands for the strong
correction. The recent measurements of the ratio~(\ref{RTauVDef})
gave $\RTau{V}=1.775 \pm 0.017$ (ALEPH Collaboration,
Ref.~\cite{ALEPH}) and $\RTau{V}=1.764 \pm 0.016$ (OPAL
Collaboration, Ref.~\cite{OPAL}).  Assuming that these data have
equal statistical weights, one arrives at the averaged value
\begin{equation}
\label{RTauVExp}
\RTau{V} = 1.769 \pm 0.017.
\end{equation}

     In the framework of the approach in hand the strong correction
in Eq.~(\ref{RTauVDef}) at the one-loop level takes the form
\begin{equation}
\label{TauCorDef}
\dQCD{(1)}=\frac{2}{\pi}\int_{0}^{M_{\tau}^2}
\left(1 - \frac{s}{M_{\tau}^2}\right)^{\!2}
\left(1 + 2 \frac{s}{M_{\tau}^2}\right) \hal{(1)}{}(s)
\frac{d s}{M_{\tau}^2},
\end{equation}
see, e.g, papers~\cite{APTTau,PRD2,Review} and references therein.
In Eq.~(\ref{TauCorDef}) $\hal{(1)}{}(s)$ is the one-loop strong
running coupling in the timelike region, and
$M_{\tau}=(1776.99_{-0.26}^{+0.29})$~MeV denotes the mass of the
$\tau$~lepton~\cite{PDG04}. As was shown in previous sections, the
mass of the $\pi$~meson entering the dispersion relation for the
\ADf~(\ref{AdlerDisp}) affects the low energy behavior of the QCD
effective charge~$\hal{}{}(s)$. Consequently, handling the
experimental data on the inclusive $\tau$~lepton decay is different
for the cases of massless and massive pion. In order to demonstrate
how the estimation of the QCD scale parameter $\Lambda$ is affected
by the nonvanishing mass of the $\pi$~meson, let us study both
instances.

     For the limit of massless pion $\mpi=0$, the one-loop strong
correction~$\dQCD{}$ to the $\RTau{V}$ ratio~(\ref{RTauVDef}) is
given by Eq.~(\ref{TauCorDef}), with $\hal{(1)}{}(s)$ being the
one-loop massless effective charge of Eq.~(\ref{AICTL}).  Although
the latter possesses the enhancement at $s \to 0$ (see
Eq.~(\ref{AICTLIR})), the resulting singularity is integrable. Then,
it is useful to represent the QCD correction in a more convenient
form
\begin{eqnarray}
\dQCD{(1)}(\MTau^2) &=&
\frac{4}{\bz} \int_{0}^{1} \left(\xi^3 - 2 \xi^2 + 2\right)
\ro{1}(\xi \mu) \, d \xi
\nonumber \\* & & +
\frac{1}{\pi} \hal{(1)}{an}(\MTau^2),
\end{eqnarray}
where $\hal{(1)}{an}(s)$ is the running coupling~(\ref{AICTL}),
$\ro{1}(\sigma)$ denotes the one-loop spectral
density~(\ref{SpDns1L}), and the notations $\xi=s/\MTau^2$ and
$\mu=\MTau^2/\Lambda^2$ are used. For the experimental data given in
Eq.~(\ref{RTauVExp}) one gets the value $\Lambda=(508\pm61)$~MeV for
the QCD scale parameter. This estimation corresponds to $\nf=2$
active quarks, and its uncertainty is due to the errors in the values
of $\RTau{V}$, $|\Vud|$, $\Sew$, and $M_{\tau}$. The relevant
behavior of the massless analytic invariant charge in the spacelike
and timelike regions (Eqs.~(\ref{AICHLKL}) and~(\ref{AICTL}),
respectively) is shown in Figure~\ref{Plot:AICTL}.

     Let us proceed now to the case of the nonvanishing pion mass.
Here, the one-loop QCD correction to the $\RTau{V}$~ratio reads as
\begin{eqnarray}
\dQCD{(1)}(\MTau^2,\mpi^2) &=& \frac{2}{\pi}\int_{0}^{M_{\tau}^2}
\left(1 - \frac{s}{M_{\tau}^2}\right)^{\!2}
\nonumber \\* && \times \!
\left(\!1 + 2 \frac{s}{M_{\tau}^2}\!\right)\! \hal{(1)}{an}(s,\mpi^2)
\frac{d s}{M_{\tau}^2},
\label{TauCorM}
\end{eqnarray}
where $\hal{(1)}{an}(s,\mpi^2)$ is the one-loop massive analytic
charge of Eq.~(\ref{AICMTL}) and $\mpi=(134.9766 \pm 0.0006)$~MeV
stands for the $\pi^0$~meson mass~\cite{PDG04}. In general, in the
framework of the analytic approach there is no need to involve the
contour integration in Eq.~(\ref{TauCorDef}), since the effective
charge $\hal{}{}(s)$, appearing in the integrand, contains no
unphysical singularities in the region~$s\ge0$. In other words, the
integration in Eq.~(\ref{TauCorDef}) can be performed in a
straightforward way.  Thus, one can cast the strong
correction~(\ref{TauCorM}) into a convenient form
\begin{eqnarray}
\dQCD{(1)}(\MTau^2, \mpi^2) &=&
\frac{4}{\bz} \int_{\xi_0}^{1}
\left(\xi^3 - 2 \xi^2 + 2\right)
\ro{1}(\xi \mu) \, d \xi
\nonumber \\* && +
\frac{1}{\pi} \hal{(1)}{an}(\MTau^2, \mpi^2),
\end{eqnarray}
where $\xi_0 = 4\mpi^2/M_{\tau}^2$ and the other notations have been
explained above. For the experimental data~(\ref{RTauVExp}) the
estimation\footnote{It is worthwhile to note here that the one-loop
perturbative analysis of the strong correction in
Eq.~(\ref{RTauVDef}) (see, e.g., Ref.~\cite{BNP}) gives the value of
the QCD scale parameter $\Lambda=(690\pm57)$~MeV for two active
quarks.} $\Lambda = (623\pm81)$~MeV has been obtained for $\nf=2$
active quarks. The uncertainty here is because of the errors of
$\RTau{V}$, $|\Vud|$, $\Sew$, $M_{\tau}$, and~$\mpi$.  The
corresponding infrared limiting value of the massive effective
charge~(\ref{AICMIR}) is $\al{(1)}{0}=1.475 \pm 0.170$.  The low
energy behavior of the analytic running coupling in the spacelike and
timelike domains (Eqs.~(\ref{AICMHL}) and~(\ref{AICMTL}),
respectively) is presented in Figure~\ref{Plot:AICM}.

     Thus, in the framework of the approach in hand it proves to be
important to take into account the mass of the $\pi$~meson in
processing the low energy QCD data. Specifically, the relative
difference between the obtained estimations of the scale parameter
$\Lambda$ for the massive and massless cases is about $20\,\%$. This
is so by virtue of the fact that the contribution to the strong
correction~(\ref{TauCorDef}) of the effects due to the pion mass
\begin{equation}
\label{DeltaTauCor}
\Delta\dQCD{(1)}(\MTau^2, \mpi^2) =
\frac{4}{\bz} \int_{0}^{\xi_0}
\left(\xi^3 - 2 \xi^2 + 2\right)
\ro{1}(\xi \mu) \, d \xi
\end{equation}
turns out to be significant. At the same time, since the scales
involved in the integral~(\ref{DeltaTauCor}) are very low, for some
models for the analytic running coupling the difference between the
limits of massive and massless $\pi$~meson may not be so sizable. For
example, in the case of the Shirkov--Solovtsov model~\cite{ShSol},
where the relevant spectral density reads as
$\rho\ind{(1)}{ss}(\sigma) = 1/(\ln^2\!\sigma+\pi^2)$, the relative
difference between the values of the QCD scale parameter, extracted
from the experimental data on the inclusive $\tau$~lepton
decay~(\ref{RTauVExp}), is about $1\,\%$, but the obtained
estimations appear to be rather large. Namely, at the one-loop level
with $\nf=2$ active quarks one gets the values $\Lambda =
(965^{+280}_{-212})$~MeV for the case of the massless $\pi$~meson,
and $\Lambda = (976^{+281}_{-213})$~MeV for the nonvanishing pion
mass.

\section{Applicability to the chiral symmetry breaking}
\label{Sect:CSB}

     Based on the study of \textit{gauge invariant} \SD equations,
Cornwall proposed a long time ago that the self-interactions of
gluons give rise to a dynamical gluon mass, while preserving at the
same time the local gauge symmetry of the
theory~\cite{Cornwall:1982zr}. This gluon ``mass'' is not a directly
measurable quantity, but has to be related with other physical
quantities, such as the glueball spectrum, the energy needed to pop
two gluons out of the vacuum, the QCD string tension, or the QCD
vacuum energy (see paper~\cite{Aguilar:2002tc} and references
therein).

     One of the main phenomenological implications of this analysis
is that the presence of the gluon mass $m_g$
saturates\footnote{Another discussion of the impact of the gluon mass
on the infrared behavior of the strong running coupling can be found,
e.g., in Ref.~\cite{DV99}.} the running of the strong coupling at low
energies.  Namely, instead of increasing indefinitely in the
infrared, as perturbation theory predicts, it ``freezes'' at a finite
value, determined by the gluon mass.  In particular, the
nonperturbative effective coupling obtained in
Ref.~\cite{Cornwall:1982zr} is given by
\begin{equation}
\label{JMC}
\alc(q^2)= \frac{4\pi}{\beta_0}
\frac{1}{\ln\!\left[z + 4M_g^2(q^2)/\Lambda^2 \right]},
\qquad z= \frac{q^2}{\Lambda^2},
\end{equation}
where $M_g(q^2)$ denotes the dynamical gluon mass
\begin{equation}
\label{runmass}
M^2_g(q^2) = m_g^2 \left[\frac{\ln\!\left(z + 4 m_g^2/\Lambda^2\right)}
{\ln\!\left(4 m_g^2/\Lambda^2\right)}\right]^{-\frac{12}{11}}.
\end{equation}
Here the nontrivial dependence of the dynamically generated gluon
mass~(\ref{runmass}) on the momentum~$q^2$ is crucial for the
renormalizability of the theory. The running coupling~(\ref{JMC}) has
the infrared finite limiting value $\alc(0) = 4\pi \left[ \bz \ln (4
m_g^2/\Lambda^2)\right]^{-1}$. It is worth noting that the above
equation makes sense only for the gluon mass satisfying $m_g
>\Lambda/2$.  For a typical values of $m_g = 500$~MeV and $\Lambda =
300$~MeV, one obtains for the case of the pure gluodynamics ($\nf=0$)
an estimation $\alc(0)\simeq0.5$. An independent analysis presented
in Ref.~\cite{Cornwall:1989gv} yields a maximum allowed value for
$\alc(0)$ of about~0.6. The incorporation of fermions into the
effective charge~\cite{Papavassiliou:1991hx} does not change the
picture qualitatively (at least for the quark masses of the order
of~$\Lambda$), resulting in an approximate expression
\begin{equation}
\label{CPEC}
\al{}{cp}(q^2) =
\frac{4\pi}{11\ln(z+\chi_g)-2\nf\ln(z+\chi_q)/3}.
\end{equation}
In this equation $\chi_g=4m_g^2/\Lambda^2$,
$\chi_q=4m_q^2/\Lambda^2$, a light quark constituent mass is $m_q =
350$~MeV~\cite{PDG04}, and $m_g = (500 \pm 100)$~MeV stands for the
gluon mass. The effective coupling of Eq.~(\ref{JMC}) was the focal
point of extensive scrutiny, and has been demonstrated to furnish an
unified description of a wide variety of the low energy QCD
data~\cite{Cristina}.

\begin{figure}[t]
\includegraphics[width=85mm]{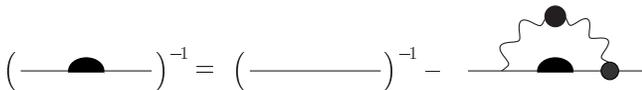}
\caption{Graphical representation of the gap
equation~(\protect\ref{gap}).}
\label{Plot:Gap}
\end{figure}

\begin{figure}[t]
\includegraphics[width=85mm]{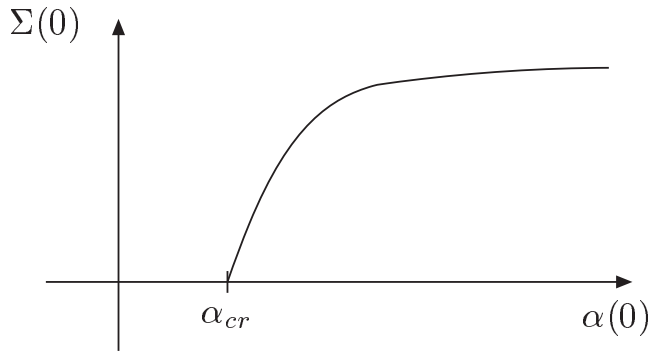}
\caption{A typical dependence of $\Sigma(0)$ on the infrared
limiting value of the QCD effective charge~$\alpha(0)$,
$\,S^{-1}(p) = A(p)\pslush + \Sigma(p)$.}
\label{Plot:CritCoupl}
\end{figure}

     In general, an important unresolved question in this context is
the incorporation of the QCD effective charge into the standard \SD
equation governing the dynamics of the quark propagator $S(p)$
\begin{equation}
\label{gap}
S^{-1}(p) = S_{0}^{-1}(p) - g^2 \int \frac{d^4k}{(2\pi)^4}
\gamma_{\mu} \, S \,\Gamma_{\nu} \,\Delta^{\mu\nu},
\end{equation}
see Figure~\ref{Plot:Gap} also. In particular, since QCD is not a
fixed point theory, the usual QED--inspired gap equation must be
modified, in order to incorporate the running charge and asymptotic
freedom. The usual way of accomplishing this eventually reduces to
the replacement $1/k^2 \to \alpha(k^2)/k^2$ in the corresponding
kernel of the gap equation, where $\alpha(k^2)$ is the QCD running
coupling. The inclusion of $\alpha(k^2)$ is essential for arriving at
an integral equation for~$S(p)$ which is well-behaved in the
ultraviolet. Indeed, the additional logarithm in the denominator of
the kernel due to the running coupling $\alpha(k^2)$ improves the
convergence of the integral.  However, since the perturbative form of
$\alpha(k^2)$ diverges at low energies as $1/\ln(k^2/\Lambda^2)$ when
$k^2 \to \Lambda^2$, some form of the infrared regularization for the
invariant charge $\alpha(k^2)$ is needed, whose details depend on the
specific assumptions one is making regarding the nonperturbative
hadron dynamics.  At this point the issue of the critical coupling
makes its appearance. Specifically, as is well-known, there is a
critical infrared limiting value of the running coupling, to be
denoted by $\al{}{cr}$, below which there are no nontrivial solutions
to the resulting gap equation, i.e., there is no chiral symmetry
breaking, see Figure~\ref{Plot:CritCoupl}. Thus, the invariant charge
$\alpha(k^2)$ employed within the gap equation must be such that
(i)~it gives rise to a nonsingular answer, (ii)~it reaches large
enough values at $k^2 \to 0$ in order to overcome~$\al{}{cr}$, and
(iii)~it does not contradict existing low-energy experimental
results.

     The incorporation of the effective charge of Eq.~(\ref{JMC})
into a gap equation has been studied for the first time in
Ref.~\cite{Haeri:1990yj}. There it was concluded that chiral symmetry
breaking solutions for $\Sigma(p)$ could be obtained only for
unnaturally small values of the gluon mass, namely $m_g/\Lambda
\simeq 0.8$. This is so because the typical value of $\al{}{cr}$
found in the standard treatment of the gap equation\footnote{The
exact value of $\al{}{cr}$ depends on the number of active flavors as
well as on the various approximations employed in deriving the gap
equations, such as the choice of gauge, or the inclusion of
gauge-technique inspired Ans\"atze for the quark-gluon vertex, but
these issues do not alter significantly our qualitative discussion.}
is $\al{}{cr} \simeq 1.2$ (see, e.g., Ref.~\cite{Atkinson}), which is
what the expression for $\alc(0)$ yields for the above ratio
of~$m_g/\Lambda$. This issue was further investigated in
Ref.~\cite{Papavassiliou:1991hx}, where a system of coupled gap and
vertex equations was considered. The upshot of this study was that no
consistent solutions to the system of integral equations could be
found, due to the fact that the allowed values for $\alpha(0)$,
dictated by the vertex equation, were significantly lower than
$\al{}{cr}$, i.e., not large enough to trigger chiral symmetry
breaking. A similar analysis was presented in
Ref.~\cite{Aguilar:2001zy}, together with several other models for
the nonperturbative QCD running coupling~\cite{Aguilar:2000kp}.

\begin{figure}[t]
\includegraphics[width=85mm]{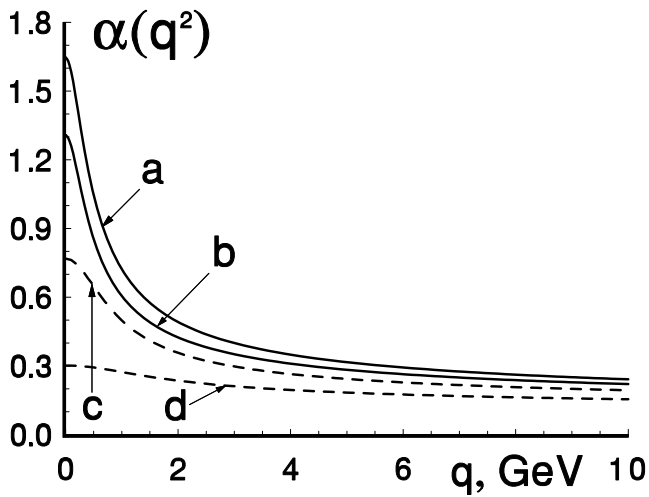}
\caption{Comparison of the massive analytic running
coupling~(\protect\ref{AICMHL}) (solid curves) with the
effective charge~(\protect\ref{CPEC}) (dashed curves).
The values of parameters are: $\nf=2$ active quarks,
$\Lambda = 704$~MeV (a), $\Lambda = 542$~MeV (b), gluon mass
$m_g=400$~MeV and $\Lambda = 350$~MeV (c), gluon mass
$m_g=600$~MeV and $\Lambda = 150$~MeV~(d).}
\label{Plot:TwoChar}
\end{figure}

     In what follows we will suggest a possible resolution of this
problem, inspired by the infrared behavior of the massive analytic
invariant charge of Eq.~(\ref{AICMHL}). The basic observation is
captured in Figure~\ref{Plot:TwoChar}. Namely, the effective charge
with a gluon mass (dashed curves) and the analytic charge (solid
curves) coincide for a large range of momenta, and they only begin to
differ appreciably in the deep infrared domain
$k^2\lesssim\Lambda^2$. In this region the analytic
charge~(\ref{AICMHL}) rises abruptly, almost doubling its size
between $k^2=\Lambda^2$ and $k^2=0$, whereas the running
coupling~(\ref{CPEC}) in the same momentum interval remains
essentially fixed to a value\footnote{The precise numerical values of
$m_g/\Lambda$ and $m_{\pi}/\Lambda$ do not change qualitatively this
picture, as long as the two scales are well separated.} of about~0.6.
A possible picture that emerges from this observation is the
following. It may be that the concept of the dynamically generated
gluon mass fails to capture all the relevant dynamics in the very
deep infrared, where confinement or other nonperturbative effects
make their appearance. At that point it could be preferable to switch
to a description in terms of the analytic charge~(\ref{AICMHL}),
which (i)~coincides with that of Cornwall in the region where the
latter furnishes a successful description of data, and (ii)~since it
overcomes the critical value $\al{}{cr}$, offers the possibility of
accounting for chiral symmetry breaking at the level of gap
equations.

\section{Conclusions}
\label{Sect:Concl}

     In this paper the effects due to the mass of the $\pi$~meson are
incorporated into the analytic approach to QCD.  The nonvanishing
pion mass gives rise to an infrared finite limiting value for the QCD
effective charge.  Besides, the latter acquires the plateau-like
behavior in the deep infrared domain of the timelike region
$0\le\sqrt{s}\le2\mpi$.  It is of a particular interest to note that
such stagnation is also predicted by a number of phenomenological
models for the strong running coupling. The developed analytic
effective charge is applied to processing the experimental data on
the inclusive $\tau$~lepton decay. The effects due to the pion mass
play a substantial role here, affecting the estimation of the QCD
scale parameter~$\Lambda$. A quantitative conclusion on the
applicability of the obtained massive running coupling to the study
of chiral symmetry breaking is drawn.

     It would be interesting to further scrutinize the developed
approach. First of all, it is of particular relevance to include the
higher order perturbative corrections in the study of the
experimental data on the inclusive $\tau$~lepton decay. Moreover, it
might also be important to incorporate the nonperturbative terms,
arising from the operator product expansion (see also
Ref.~\cite{Maxwell}) and from the so-called nonlocal chiral quark
model~\cite{Dorokhov}, into the \ADf.  In addition, a detailed study
of the gap equation, with the analytic charged plugged into it, is
needed in order to verify if indeed one encounters nontrivial
solutions, whose size is phenomenologically relevant. Specifically,
one should check by making use of, e.g., the Pagels--Stokar
method~\cite{Pagels:1979hd}, whether the solutions obtained for
$\Sigma(p)$ can reproduce a reasonable value of the pion-decay
constant $f_{\pi}$.  It is also interesting to apply the developed
model to the study of the pion electromagnetic form factor
$F_{\pi}(qý)$ (see paper~\cite{Stefanis} and references therein). At
the same time, a crucial point to explore is whether the massive
analytic effective charge satisfies a variety of phenomenological
constraints, imposed by the low energy experimental data on the
infrared behavior of the QCD running coupling, see, e.g.,
papers~\cite{Doksh,Cristina} and references therein.

\begin{acknowledgments}
     The authors thank Professors D.V.\ Shirkov, A.C.\ Aguilar, A.E.\
Dorokhov, I.L.\ Solovtsov, and N.G.\ Stefanis for the stimulating
discussions and useful comments.  This work was supported by grants
SB2003-0065 of the Spanish Ministry of Education, CICYT
FPA20002-00612, RFBR (Nos.\ 02-01-00601 and 04-02-81025), and
NS-2339.2003.2.
\end{acknowledgments}


\begin{thebibliography}{99}

\bibitem{AF} D.J.~Gross and F.~Wilczek,
          Phys.\ Rev.\ Lett.\ \textbf{30}, 1343 (1973);
         H.D.~Politzer, \textit{ibid.} \textbf{30}, 1346 (1973);
         G.~'t~Hooft, report at the Conference on Yang--Mills
          Fields (Marseille, France, 1972).

\bibitem{RG} E.C.G.~Stuckelberg and A.~Petermann, Helv.\ Phys.\ Acta
         \textbf{24}, 317 (1951); \textbf{26}, 499 (1953); M.~Gell--Mann and
         F.E.~Low, Phys.\ Rev.\ \textbf{95}, 1300 (1954);
         N.N.~Bogoliubov and D.V.~Shirkov, Dokl.\ Akad.\
         Nauk SSSR \textbf{103}, 203 (1955); \textbf{103}, 391 (1955);
         Nuovo Cimento \textbf{3}, 845 (1956).

\bibitem{BgSh} N.N.~Bogoliubov and D.V.~Shirkov, \textit{Introduction to
         the Theory of Quantized Fields} (Interscience, New York, 1980).

\bibitem{Peter} W.~Fischler, Nucl.\ Phys.\ B \textbf{129}, 157 (1977);
         T.~Appelquist, M.~Dine, and I.J.~Muzinich, Phys.\ Lett.\
         B \textbf{69}, 231 (1977); Phys.\ Rev.\ D \textbf{17}, 2074 (1978);
         M.~Peter, Phys.\ Rev.\ Lett.\ \textbf{78}, 602 (1997);
         Nucl.\ Phys.\ B \textbf{501}, 471 (1997).

\bibitem{LatQCD} G.S.~Bali \textit{et al.} (SESAM and T$\chi$L
          Collaborations), Phys.\ Rev.\ D \textbf{62}, 054503 (2000);
         G.S.~Bali, Phys.\ Rep.\ \textbf{343}, 1 (2001);
         C.~Bernard \textit{et al.}, Phys.\ Rev.\ D \textbf{62}, 034503
          (2000);
         S.~Necco and R.~Sommer, Nucl.\ Phys.\ B \textbf{622}, 328
          (2002);
         T.T.~Takahashi, H.~Suganuma, Y.~Nemoto, and H.~Matsufuru,
          Phys.\ Rev.\ D \textbf{65}, 114509 (2002);
         S.~Aoki \textit{et al.} (JLQCD Collaboration), \textit{ibid.}
          \textbf{68}, 054502 (2003).

\bibitem{Nest} B.M.~Barbashov and V.V.~Nesterenko, \textit{Introduction
         to the Relativistic String Theory} (World Scientific, Singapore,
         1990).

\bibitem{VScheme} W.~Celmaster and F.S.~Henyey, Phys.\ Rev.\ D \textbf{18},
          1688 (1978);
         D.B.~Lichtenberg and J.G.~Wills, Nuovo Cimento A \textbf{47},
          483 (1978);
         R.~Levine and Y.~Tomozawa, Phys.\ Rev.\ D \textbf{19}, 1572 (1979);
         J.L.~Richardson, Phys.\ Lett.\ B \textbf{82}, 272 (1979);
         W.~Buchmuller, G.~Grunberg, and S.-H.H.~Tye, Phys.\ Rev.\ Lett.\
          \textbf{45}, 103 (1980).

\bibitem{HadrSpectr} W.~Lucha, F.F.~Schoberl, and D.~Gromes,
          Phys.\ Rept.\ \textbf{200}, 127 (1991);
         N.~Brambilla and A.~Vairo, arXiv:hep-ph/9904330;
         V.V.~Kiselev, A.E.~Kovalsky, and A.I.~Onishchenko,
          Phys.\ Rev.\ D \textbf{64}, 054009 (2001).

\bibitem{IScheme} A.~Ringwald and F.~Schrempp, Phys.\ Lett.\ B \textbf{459},
         249 (1999).

\bibitem{UKQCD} D.A.~Smith and M.J.~Teper (UKQCD Collaboration),
         Phys.\ Rev.\ D \textbf{58}, 014505 (1998).

\bibitem{PRD1} A.V.~Nesterenko, Phys.\ Rev.\ D \textbf{62}, 094028 (2000).

\bibitem{PRD2} A.V.~Nesterenko, Phys.\ Rev.\ D \textbf{64}, 116009 (2001).

\bibitem{Schrempp} F.~Schrempp, J. Phys.\ G \textbf{28}, 915 (2002).

\bibitem{Review} A.V.~Nesterenko, Int.\ J.\ Mod.\ Phys.\ A \textbf{18},
         5475 (2003).

\bibitem{OPT} P.M.~Stevenson, Phys.\ Rev.\ D \textbf{23}, 2916 (1981).

\bibitem{OPTEpEm} A.C.~Mattingly and P.M.~Stevenson, Phys.\ Rev.\ D
         \textbf{49}, 437 (1994).

\bibitem{EffChar} G.~Grunberg, Phys.\ Rev.\ D \textbf{29}, 2315 (1984).

\bibitem{BLM} S.J.~Brodsky, G.P.~Lepage, and P.B.~Mackenzie,
         Phys.\ Rev.\ D \textbf{28}, 228 (1983).

\bibitem{Fischer} S.~Ciulli and J.~Fischer, Nucl.\ Phys.\ \textbf{24},
         465 (1961); J.~Fischer, Fortschr.\ Phys.\ \textbf{42}, 665
	 (1994); I.~Caprini and J.~Fischer, Phys.\ Rev.\ D \textbf{60},
	 054014 (1999); \textbf{62}, 054007 (2000); \textbf{68}, 114010
	 (2003).

\bibitem{Elias1} V.~Elias, D.G.C.~McKeon, and T.G.~Steele, Phys.\
         Rev.\ D \textbf{69}, 045015 (2004); Int.\ J.\ Mod.\ Phys.\
	 A \textbf{18}, 3417 (2003); V.~Elias, arXiv:hep-ph/0305187.

\bibitem{Elias2} M.R.~Ahmady \textit{et al.}, Nucl.\ Phys.\ B \textbf{655},
         221 (2003); Phys.\ Rev.\ D \textbf{66}, 014010 (2002); \textbf{67},
	 034017 (2003).

\bibitem{AQED} P.J.~Redmond, Phys.\ Rev.\ \textbf{112}, 1404 (1958);
         P.J.~Redmond and J.L.~Uretsky, Phys.\ Rev.\ Lett.\
         \textbf{1}, 147 (1958); N.N.~Bogoliubov, A.A.~Logunov, and
         D.V.~Shirkov, Zh.\ Eksp.\ Teor.\ Fiz.\ \textbf{37}, 805 (1959)
         [Sov.\ Phys.\ JETP \textbf{37}, 574 (1960)].

\bibitem{Bjorken} J.D.~Bjorken, report SLAC-PUB-5103 (1989).

\bibitem{DMW} Y.L.~Dokshitzer, G.~Marchesini, and B.R.~Webber,
         Nucl.\ Phys.\ B \textbf{469}, 93 (1996).

\bibitem{ShSol} D.V.~Shirkov and I.L.~Solovtsov,
         JINR Rapid Comm.\ \textbf{2}, 5 (1996);
         Phys.\ Rev.\ Lett.\ \textbf{79}, 1209 (1997).

\bibitem{APTRev} D.V.~Shirkov, Teor.\ Mat.\ Fiz.\ \textbf{119}, 55 (1999)
         [Theor.\ Math.\ Phys.\ \textbf{119}, 438 (1999)];
         I.L.\ Solovtsov and D.V.\ Shirkov,
         Teor.\ Mat.\ Fiz.\ \textbf{120}, 482 (1999)
         [Theor.\ Math.\ Phys.\ \textbf{120}, 1220 (1999)].

\bibitem{QCD03} A.V.~Nesterenko, Nucl.\ Phys.\ B (Proc.\ Suppl.)
         \textbf{133}, 59 (2004).

\bibitem{MS97} K.A.~Milton and I.L.~Solovtsov, Phys.\ Rev.\ D \textbf{55},
         5295 (1997); \textbf{59}, 107701 (1999).

\bibitem{Cornwall:1982zr}
J.M.~Cornwall,
Phys.\ Rev.\ D \textbf{26}, 1453 (1982).

\bibitem{DISRG} A.~Peterman, Phys.\ Rep.\ \textbf{53}, 157 (1979);
         A.J.~Buras, Rev.\ Mod.\ Phys.\ \textbf{52}, 199 (1980).

\bibitem{PenRos} M.R.~Pennington and G.G.~Ross, Phys.\ Lett.\ B
         \textbf{102}, 167 (1981).

\bibitem{Adler} S.L.~Adler, Phys.\ Rev.\ D \textbf{10}, 3714 (1974).

\bibitem{Rad82} A.V.~Radyushkin, Joint Institute for Nuclear Research
         report No.\ 2--82--159 (1982); JINR Rapid Comm.\ \textbf{4}, 9
         (1996); arXiv:hep-ph/9907228.

\bibitem{KrPi82}  N.V.~Krasnikov and A.A.~Pivovarov, Phys.\ Lett.\ B
         \textbf{116}, 168 (1982).

\bibitem{Repem} T.~Appelquist and H.~Georgi, Phys.\ Rev.\ D \textbf{8},
         4000 (1973); A.~Zee, \textit{ibid.} D \textbf{8}, 4038 (1973).

\bibitem{GKL91} S.G.~Gorishny, A.L.~Kataev, and S.A.~Larin, Phys.\
         Lett.\ B \textbf{259}, 144 (1991).

\bibitem{SurSa91} L.R.~Surguladze and M.A.~Samuel, Phys.\ Rev.\ Lett.\
         \textbf{66}, 560 (1991); textbf{66}, 2416(E) (1991).

\bibitem{DV01} D.V.~Shirkov, Eur.\ Phys.\ J.\ C \textbf{22}, 331 (2001);
         Teor.\ Mat.\ Fiz.\ \textbf{127}, 3 (2001) [Theor.\ Math.\ Phys.\
         \textbf{127}, 409 (2001)].

\bibitem{APTTau} K.A.~Milton, I.L.~Solovtsov, and O.P.~Solovtsova,
         Phys.\ Rev.\ D \textbf{64}, 016005 (2001); D \textbf{65}, 076009
         (2002); K.A.\ Milton, I.L.\ Solovtsov, O.P.\ Solovtsova, and
         V.I.\ Yasnov, Eur.\ Phys.\ J.\ C \textbf{14}, 495 (2000).

\bibitem{JLD} R.~Jost and H.~Lehmann, Nuovo Cimento \textbf{5}, 1598 (1957);
         F.J.~Dyson, Phys.\ Rev.\ \textbf{110}, 1460 (1958);
	 N.N.~Bogoliubov, V.S.~Vladimirov, and A.N.~Tavkhelidze, Theor.\
	 Math.\ Phys.\ \textbf{12}, 305 (1972).

\bibitem{MPLA2} A.V.~Nesterenko and I.L.~Solovtsov, Mod.\ Phys.\ Lett.\
         A \textbf{16}, 2517 (2001).

\bibitem{DV02} D.V.~Shirkov, Teor.\ Mat.\ Fiz.\ \textbf{132}, 484 (2002)
         [Theor.\ Math.\ Phys.\ \textbf{132}, 1309 (2002)].

\bibitem{DV04} D.V.~Shirkov, in \textit{Proceedings of the Eleventh
         International
         QCD Conference (5--10 July 2004, Montpellier, France)} (to be
	 published); arXiv:hep-ph/0408272.

\bibitem{Vr} A.V.~Nesterenko, Int.\ J.\ Mod.\ Phys.\ A \textbf{19}, 3471
        (2004).

\bibitem{MPLA1} A.V.~Nesterenko, Mod.\ Phys.\ Lett.\ A \textbf{15}, 2401
         (2000).

\bibitem{Schwinger} J.~Schwinger, Proc.\ Natl.\ Acad.\ Sci.\ USA
         \textbf{71}, 3024 (1974); \textbf{71}, 5047 (1974).

\bibitem{Milton} K.A.~Milton, Phys.\ Rev.\ D \textbf{10}, 4247 (1974).

\bibitem{NPQCD04} A.V.~Nesterenko and J.~Papavassiliou, in
         \textit{Proceedings
         of the Eighth Workshop on Nonperturbative QCD (7--11 June 2004,
         Paris, France)} (to be published); arXiv:hep-ph/0409220.

\bibitem{QCD04} A.V.~Nesterenko and J.~Papavassiliou, in
         \textit{Proceedings
         of the Eleventh International QCD Conference (5--10 July 2004,
         Montpellier, France)} (to be published); arXiv:hep-ph/0410072.

\bibitem{AS} M.~Abramowitz and I.A.~Stegun (Eds.), \textit{Handbook of
         Mathematical Functions} (Dover, New York, 1972).

\bibitem{Hig84} K.~Higashijima, Phys.\ Rev.\ D \textbf{29}, 1228 (1984).

\bibitem{BDP} E.~Braaten, Phys.\ Rev.\ Lett.\ \textbf{60}, 1606 (1988);
         Phys.\ Rev.\ D \textbf{39}, 1458 (1989); F.~Le~Diberder and
         A.~Pich, Phys.\ Lett.\ B \textbf{286}, 147 (1992).

\bibitem{BNP} E.~Braaten, S.~Narison, and A.~Pich, Nucl.\ Phys.\ B
         \textbf{373}, 581 (1992).

\bibitem{PDG04} S.~Eidelman \textit{et al.} (Particle Data Group),
         Phys.\ Lett.\ B \textbf{592}, 1 (2004).

\bibitem{EWF} W.J.~Marciano and A.~Sirlin, Phys.\ Rev.\ Lett.\
         \textbf{61}, 1815 (1988); \textbf{56}, 22 (1986); E.~Braaten and
         C.S.~Li, Phys.\ Rev.\ D \textbf{42}, 3888 (1990).

\bibitem{ALEPH} R.~Barate \textit{et al.} (ALEPH Collaboration),
         Eur.\ Phys.\ J.\ C \textbf{4}, 409 (1998).

\bibitem{OPAL} K.~Ackerstaff \textit{et al.} (OPAL Collaboration),
         Eur.\ Phys.\ J.\ C \textbf{7}, 571 (1999).

\bibitem{Aguilar:2002tc}
A.C.~Aguilar, A.A.~Natale, and P.S.~Rodrigues da Silva,
Phys.\ Rev.\ Lett.\ \textbf{90}, 152001 (2003).

\bibitem{DV99} D.V.~Shirkov, Phys.\ Atom.\ Nucl.\ \textbf{62},
         1928 (1999).

\bibitem{Cornwall:1989gv}
J.M.~Cornwall and J.~Papavassiliou,
Phys.\ Rev.\ D \textbf{40}, 3474 (1989).

\bibitem{Papavassiliou:1991hx}
J.~Papavassiliou and J.M.~Cornwall,
Phys.\ Rev.\ D \textbf{44}, 1285 (1991).

\bibitem{Cristina} A.C.~Aguilar, A.~Mihara, and A.A.~Natale,
         Int.\ J.\ Mod.\ Phys.\ A \textbf{19}, 249 (2004).

\bibitem{Haeri:1990yj}
B.~Haeri and M.B.~Haeri,
Phys.\ Rev.\ D \textbf{43}, 3732 (1991).

\bibitem{Atkinson} D.~Atkinson, P.W.~Johnson, and K.~Stam,
     Phys.\ Rev.\ D \textbf{37}, 2996 (1988).

\bibitem{Aguilar:2001zy}
A.C.~Aguilar, A.~Mihara, and A.A.~Natale,
Phys.\ Rev.\ D \textbf{65}, 054011 (2002).

\bibitem{Aguilar:2000kp}
A.C.~Aguilar, A.A.~Natale, and R.~Rosenfeld,
Phys.\ Rev.\ D \textbf{62}, 094014 (2000).

\bibitem{Maxwell} D.M.~Howe and C.J.~Maxwell, Phys.\ Rev.\
         D \textbf{70}, 014002 (2004).

\bibitem{Dorokhov} A.E.~Dorokhov and W.~Broniowski, Eur.\ Phys.\ J.\
         C \textbf{32}, 79 (2003); I.V.\ Anikin, A.E.\ Dorokhov, and
	 L.\ Tomio, Phys.\ Part.\ Nucl.\ \textbf{31}, 509 (2000) [Fiz.\
	 Elem.\ Chast.\ Atom.\ Yadra \textbf{31}, 1023 (2000)];
	 A.E.~Dorokhov, arXiv:hep-ph/0405153.

\bibitem{Pagels:1979hd}
H.~Pagels and S.~Stokar,
Phys.\ Rev.\ D \textbf{20}, 2947 (1979).

\bibitem{Stefanis} A.P.~Bakulev, K.~Passek-Kumericki, W.~Schroers, and
         N.G.~Stefanis, Phys.\ Rev.\ D \textbf{70}, 033014 (2004);
	 N.G.~Stefanis, arXiv:hep-ph/0410245.

\bibitem{Doksh} Yu.L.~Dokshitzer and B.R.~Webber, Phys.\ Lett.\
         B \textbf{352}, 451 (1995); Yu.L.\ Dokshitzer, V.A.\ Khoze, and
	 S.I.\ Troyan, Phys.\ Rev.\ D \textbf{53}, 89 (1996).

\end{thebibliography}
\end{document}